# STRUCTURE AND BONDING OF SECOND-ROW HYDRIDES

S. M. Blinder*
Wolfram Research Inc., Champaign IL, 61820
and
University of Michigan, Ann Arbor, MI 48109-1055

*E-mail: sblinder@wolfram.com

The simplest hydrides of boron, carbon, nitrogen and oxygen provide a elementary picture of atomic orbitals, hybridization and chemical bonding, which can be very instructive for beginning chemistry students.[1] The valence shells of the free atoms of B, C, N and O in their ground states have the electron configurations $2s^2 2p$, $2s^2 2p^2$, $2s^2 2p^3$, $2s^2 2p^4$, respectively, (apart from their $1s^2$ inner shells). The three degenerate $2p$ orbitals are singly occupied, except for O, in which one of the $2p$ orbitals must double up. These configurations are represented graphically in Figure 1, with the electrons, shown as white dots, occupying the orbitals, whose geometrical forms are drawn schematically.

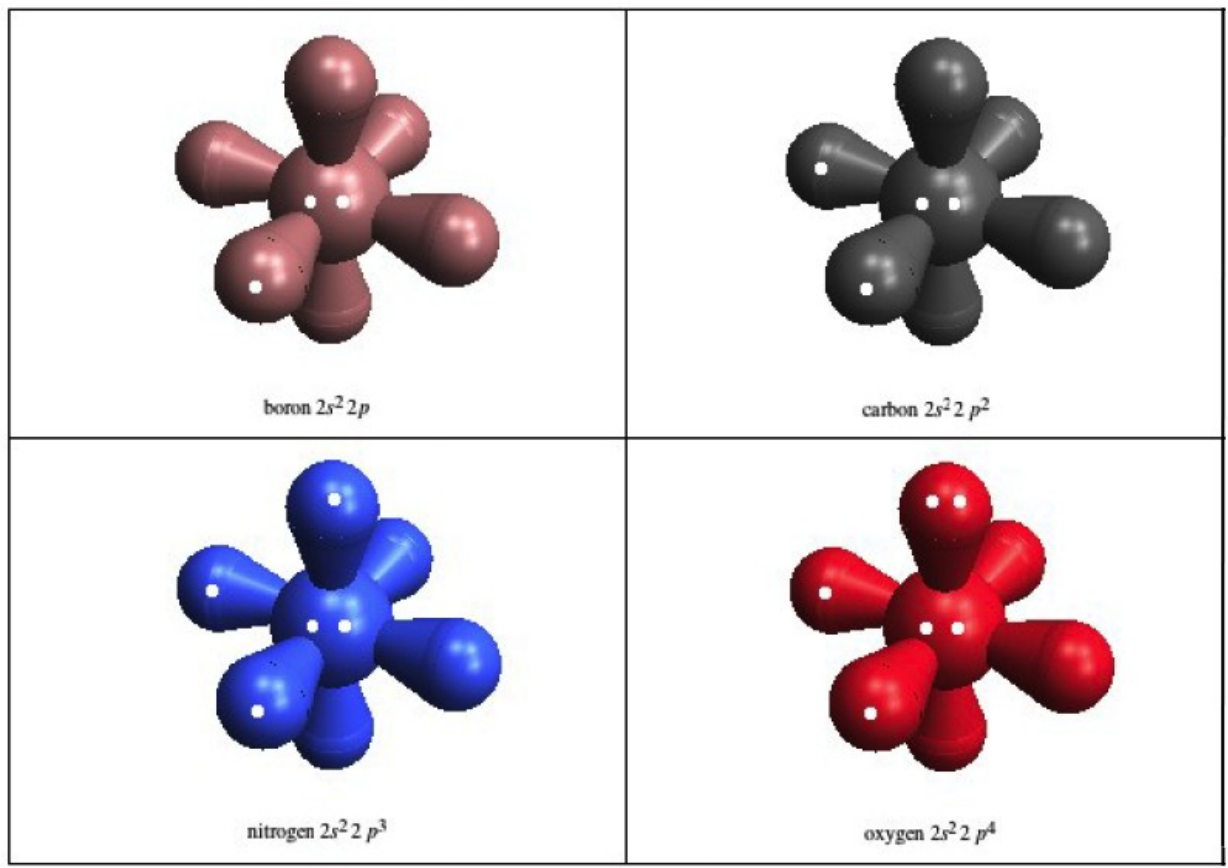

Figure 1. Ground state electron configurations of boron, carbon, nitrogen and oxygen atoms.

Carbon, with its two unpaired electrons, appears to be a naturally divalent atom, and indeed the compound $CH_2$ can exist in the gas phase. But much more stable compounds can be formed if carbon invests a relatively small amount of energy to excite one of its 2s-electrons to the remaining unoccupied 2p orbital, and thereby becomes quadrivalent, thus recouping the 2s-2p excitation energy in the formation of two additional chemical bonds. A further transformation, first suggested by Linus Pauling,[2] is the linear combination of the nearly degenerate 2s and three 2p orbitals into four identically-shaped hybrid orbitals, directed toward the corners of a tetrahedron. These are called $sp^3$-hybrid orbitals, which can be designated $t_1$, $t_2$, $t_3$, and $t_4$. The carbon atom with the four singly-occupied tetrahedral hybrids is now in what can be designated as its "valence state," a construct introduced independently by J. H. Van Vleck[3] and W. E. Moffitt,[4] as the conceptual precursor of bond formation to hydrogen atoms (or other elements). The quadrivalent valence state of carbon atom, along with the four hydrogen atoms ready to form bonds, is represented in Figure 2.

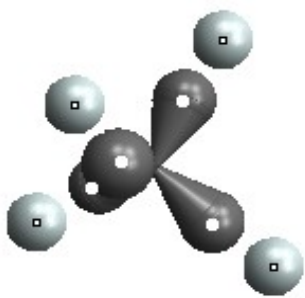

Figure 2. Valence state of carbon atom in methane.

Indeed, the methane molecule $CH_4$, formed by covalent bonding with four hydrogen atoms has a tetrahedral shape with identical angles of 109.5° between each pair of C-H bonds, as shown in the familiar ball and stick model of methane in Figure 3.

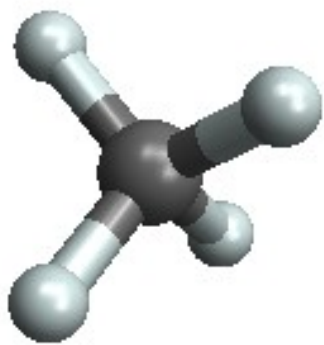

Figure 3. Molecular plot of methane, $CH_4$.

Nitrogen and oxygen atoms also tend to produce tetrahedral hybrids. Except when four identical atoms bond to the central atom, the hybrids are slightly distorted from a perfect tetrahedral shape. This picture can be subsumed by the VSEPR model of chemical bonding, in which bonds and lone pairs of electrons adopt a configuration determined by their maximized repulsions.

Nitrogen expresses its natural trivalence to form the ammonia molecule $NH_3$. It still has an approximately tetrahedral structure with a lone pair of electrons occupying one of the vertices, The N-H bond angles are reduced to 107.8º because the lone pair repels the N-H bonds. Adding an additional proton produces the ammonium ion $NH_4^+$, which is again a regular tetrahedron. Figure 4 shows the valence states which are precursors to $NH_3$ and $NH_4^+$,

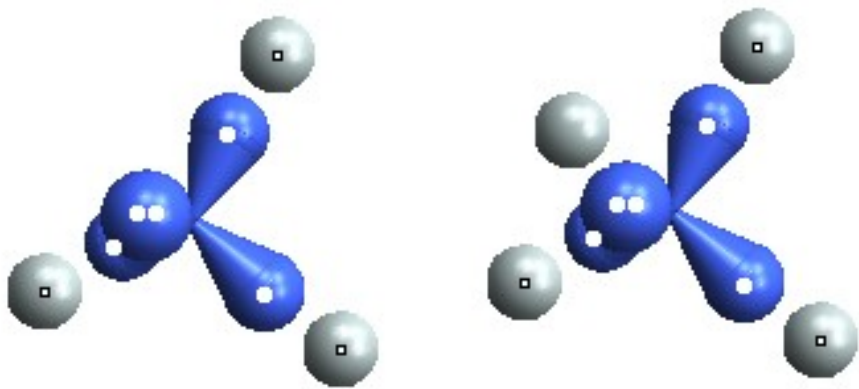

Figure 4. Valence states of nitrogen atom in $NH_3$ and $NH_4^+$.

resulting in the ammonia molecule and the ammonium ion, Figure 5.

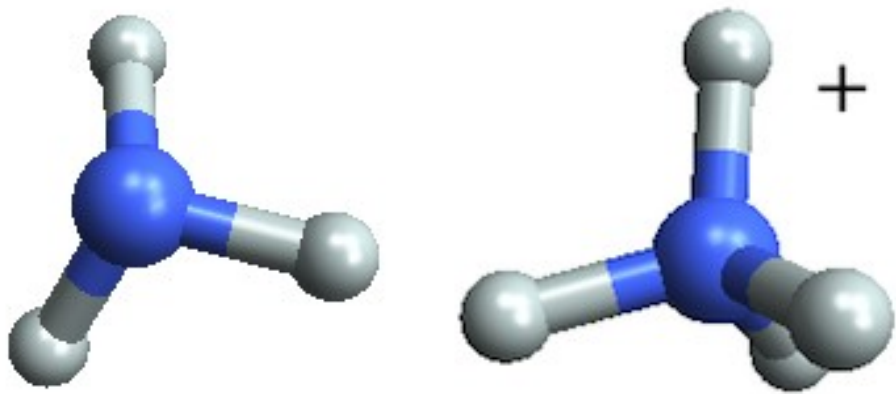

Figure 5. Ammonia molecule $NH_3$ and ammonium ion $NH_4^+$.

The best-known compound of oxygen, and an essential for life, is, of course, water $H_2O$, which forms two bonds to hydrogen atoms in addition to two lone pairs. The H-O-H angle is reduced to 104.5º by repulsion of the lone pairs. A principal component of acids is the hydronium ion $H_3O^+$, with a structure analogous to ammonia. Figures 6 and 7 show the relevant valence states and molecular structures. When water is involved in hydrogen bonding, the oxygen can momentarily be surrounded by four hydrogen atoms or ions in a tetrahedral configuration. This arrangement can occur in the crystal structure of ice.

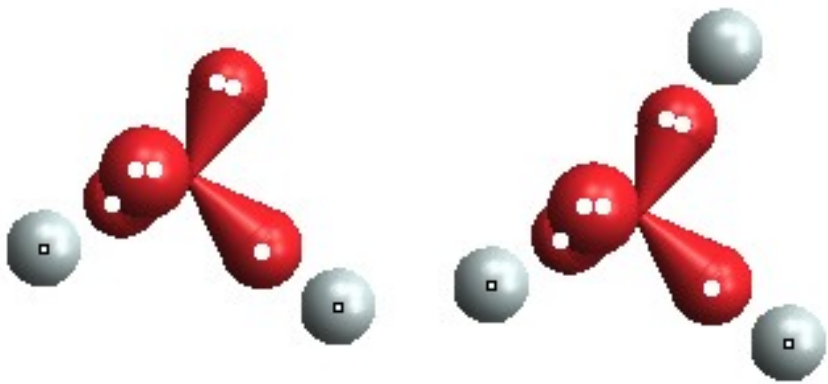

Figure 6. Valence states of oxygen in water and hydronium ion.

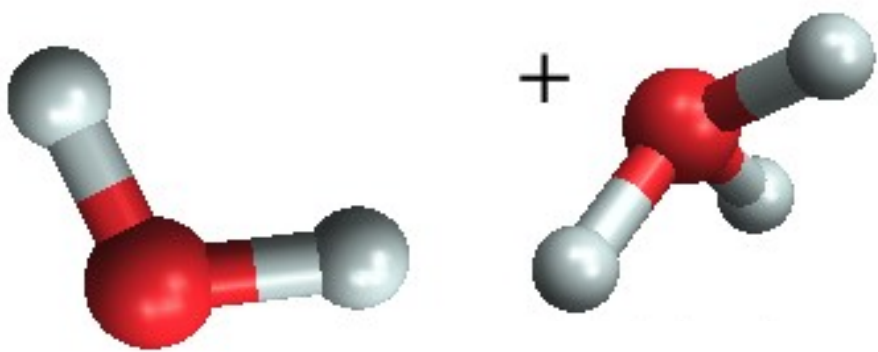

Figure 7. Molecular structures of water molecule $H_2O$ and hydronium ion $H_3O^+$.

We have saved for last, the hydrides of boron. Compounds of boron exhibit the behavior of "electron-deficient" species. This means that there are not enough electrons to permit the formation of conventional 2-electron bonds. The simplest hydride, borane $BH_3$, which can be expected to have a planar triangular structure, utilizes trigonal $sp^2$-hybrid atomic orbitals (which can be designated $tr_1$, $tr_2$, $tr_3$). It is an extremely unstable compound, however, and spontaneously dimerizes to form diborane, $B_2H_6$. (The fluorine analog $BF_3$ is a stable molecule, secured by the larger electronegativity of fluorine.) Figure 8 shows the valence state and molecular structure of $BH_3$.

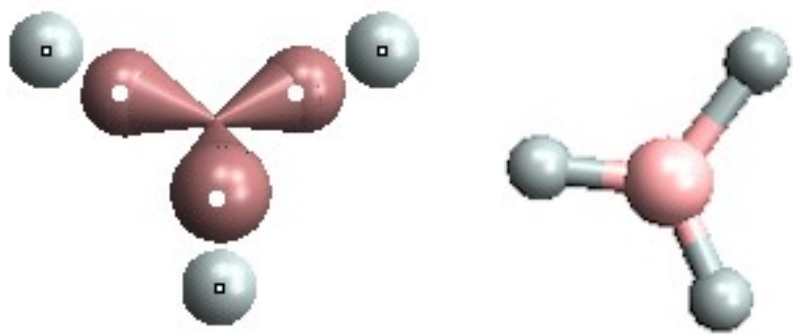

Figure 8. Trigonal valence state of boron and molecular structure of $BH_3$.

For the more common valence state of boron, one can imagine a tetrahedron with one empty orbital. This can also be pictured as a resonance hybrid in which the 3 electrons are distributed among the 4 tetrahedral lobes (with an average of 3/4 of an electron per orbital). This works to account for the tetrahedral structure of the borohydride ion $BH_4^-$, which requires combination with three hydrogen atoms, plus a hydride ion $H^-$. The valence state and the structure of the borohydride are shown in Figure 9.

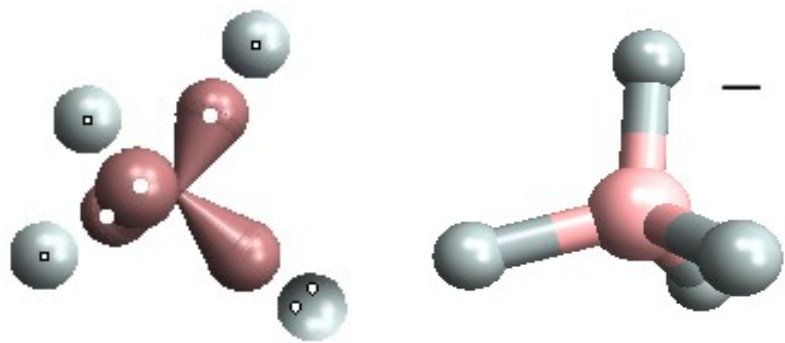

Figure 9. Tetrahedral valence state of boron and structure of $BH_4^-$.

A long controversial problem in chemistry was the structure of diborane $B_2H_6$. We propose the following picture, beginning with two boron atoms in juxtaposition occupying their hypothetical tetrahedral valence states, as shown in the above graphic. Four hydrogen atoms can be added "normally" to the ends of the molecule. The four remaining valence orbitals on the two boron atoms, having only two electrons between then can then be imagined to form four one-electron bonds with two hydrogen atoms, as shown in Figure 10.

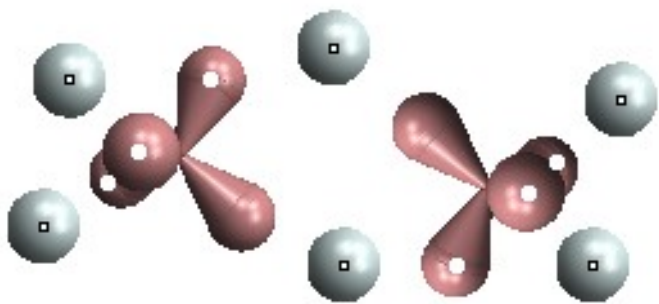

Figure 10. Proposed valence state for diborane formation.

This produces the unique bridged structure of diborane, shown in Figure 11.

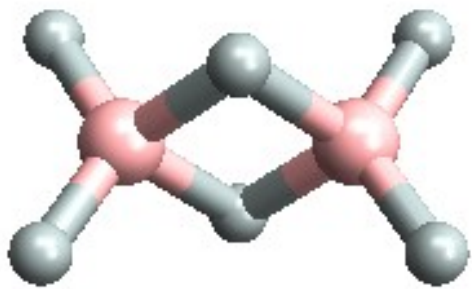

Figure 11. Structure of diborane $B_2H_6$.

One of the first to propose such a bridged structure was H. C. Longuet-Higgins.[5] The structure has since has been amply verified experimentally.[6] One-electron bonds are known to exist, for example in the $H_2^+$ molecule-ion. Alternatively, the two B-H-B bridges can be classified as 3-center, 2-electron bonds. Heavier boranes (hydrides of boron), such as $B_4H_{10}$, $B_5H_9$, etc., make extensive use of the theme of B-H-B and B-B-B bridge bonds.[7]

The graphics in this article, as well as relevant chemical data, made use of *Mathematica*™ software. I would also like to thank my colleagues at Wolfram Research for their invaluable assistance.

References:

(1) S. M. Blinder, *Introduction to Quantum Mechanics*, Elsevier: Amsterdam, 2004, pp. 142-143.
(2) L. Pauling, *The Nature of the Chemical Bond*, Cornell University Press: Ithaca, NY, 1948.
(3) J. H. Van Vleck, On the Theory of the Structure of $CH_4$ and Related Molecules: Part III, *J. Chem. Phys.* 1934 **2**, 20-30.
(4) W. Moffitt, Atomic Valence States and Chemical Binding, *Rep. Prog. Phys.* 1954, **17**, 173-200.
(5) H. C. Longuet-Higgins and R. P. Bell, The Structure of the Boron Hydrides, *J. Chem. Soc. (Resumed)* 1943, 250-255.
(6) K. Hedberg and V. Schomaker, A Reinvestigation of the Structures of Diborane and Ethane by Electron Diffraction, *Journal of the American Chemical Society* 1951, **73**, 1482–1487.
(7) See, for example, F. A. Cotton and G. Wilkinson, *Advanced Inorganic Chemistry*, Wiley Interscience: New York, 1980, pp. 303ff.